\pdfoutput=1
\documentclass[sigconf]{acmart}

\usepackage{amsmath,amsfonts}
\usepackage{algorithmic}
\usepackage{graphicx}
\usepackage{textcomp}
\usepackage{xcolor}
\usepackage{soulpos}
\usepackage{graphicx}
\usepackage{multirow}
\usepackage{listings}
\usepackage{subcaption}
\usepackage[frozencache,cachedir=.]{minted}
\usepackage{ragged2e}
\usepackage{multirow}
\usepackage{color,soul}
\def\BibTeX{{\rm B\kern-.05em{\sc i\kern-.025em b}\kern-.08em
    T\kern-.1667em\lower.7ex\hbox{E}\kern-.125emX}}
\begin{document}

\title{Efficiently Executing High-throughput Lightweight LLM Inference Applications on Heterogeneous Opportunistic GPU \\Clusters with Pervasive Context Management}

\author{Thanh Son Phung, Douglas Thain}
\email{{tphung, dthain}@nd.edu}

\affiliation{%
  \institution{University of Notre Dame}
  \city{Notre Dame}
  \state{IN}
  \country{USA}
}

\begin{abstract}

The rise of Generative AI introduces a new class of HPC workloads that integrates lightweight LLMs with traditional high-throughput applications to accelerate scientific discovery. The current design of HPC clusters is inadequate to support this new class however, either incurring long wait times on static batch queues or repeatedly paying expensive LLM startup costs upon resource preemption. To circumvent both the long queues and high startup costs, we propose to "decouple" the LLM initialization context from the actual LLM inferences, and retain the context in GPUs until it is no longer needed, a technique we term "Pervasive Context Management". We transform a fact verification application to enable this technique, allowing it to reduce its execution time by 72.1\% (from 3 hours to 48 minutes) using the same amount of GPUs, and scale opportunistically on 32.8\% of all GPUs in the cluster and further reduce the execution time to 13 minutes.

\end{abstract}

\maketitle


\section{Introduction}
\label{sec:intro}
\textbf{Background}.
Large Language Models (LLMs)\cite{achiam2023gpt, team2024gemini, anthropic_claude_3} have emerged as a transformative technology, demonstrating a remarkable capacity for discerning intricate patterns within vast datasets, and promise to be exceptionally powerful tools across diverse scientific domains. Consequently, a \textbf{new class of HPC workloads} is on the rise that integrates lightweight LLM inferencing (typically with \textit{billions} of parameters) into traditional high-throughput applications to accelerate the pace of scientific discovery. This trend, exemplified by LLM-backed advancements in protein folding\cite{wohlwendminifold,sunliteformer,shah2024energy, vieira2024scaling} and distributed AI-driven scientific computing frameworks\cite{ward2021colmena,fan2024workflowllm,gao2025strategic}, introduces a novel and challenging workload class to High-Performance Computing (HPC) clusters.

Currently, two primary allocation models exist for executing these LLM-integrated applications on GPU clusters. In the \underline{\textit{static}} \underline{\textit{allocation model}}, an application requests an exclusive and fixed-size batch of GPUs for a given period of time. This approach ensures predictable, maximal performance by eliminating resource contention issues during its execution. However, \textbf{its practicality is diminishing}: the combination of surging demand for GPU resources and high GPU market prices\cite{Deloitte_2025,pilz2025trends,kachris2025survey} has led to a huge number of GPU-accelerated applications oversubscribing a much smaller number of GPUs. This oversubscription results in heavily contended job queues, forcing applications to have significant wait times\cite{luo2024scheduling,liang2024resource, ding2023mirage}. Furthermore, the rigidity of static allocations can lead to cluster-wide resource under-utilization due to resource fragmentation, wasting idle GPU cycles\cite{jeon2018multi,xiao2020antman,jeon2019analysis}.

\begin{figure}[t]
    \includegraphics[width=\columnwidth]{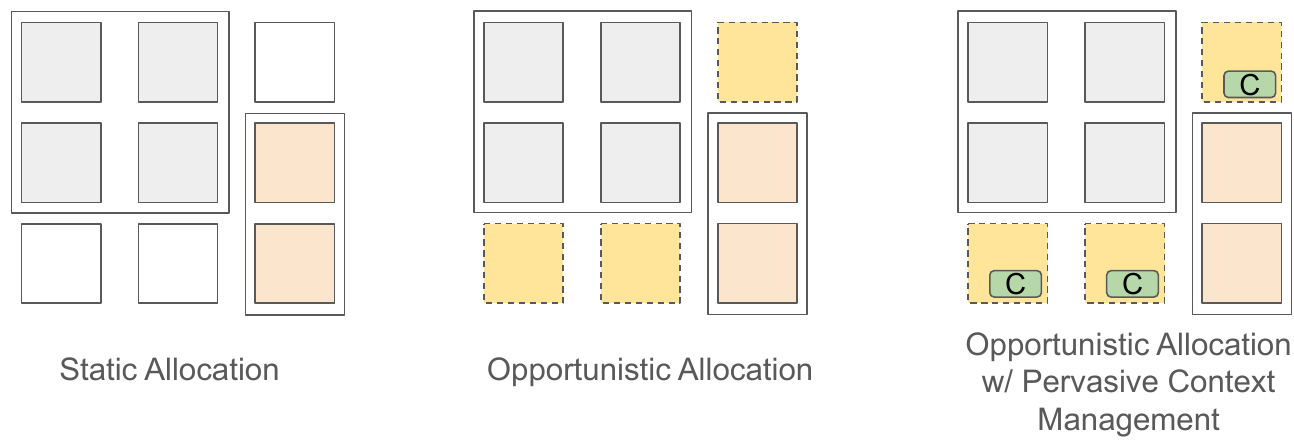}
    \caption{Different Resource Allocation Models}
    \justifying
    \noindent
    \textit{(Left) Static allocation model exclusively allocates a block of GPUs to an application, leading to long wait times on batch queues and cluster-wide resource under-utilization. (Middle) Opportunistic allocation model allows applications to run on transiently available GPUs, but incurs a high startup penalty from LLM context initialization. (Right) Our proposed model with Pervasive Context Management: LLM contexts - denoted as C - are decoupled from inferences and deliberately persisted on GPUs to alleviate the high startup penalty.}
\end{figure}

Instead of relying on fixed-size exclusive batches of GPUs, the \underline{\textit{opportunistic allocation model}} leverages dynamic resources in which GPUs can join and leave the application's resource pool at any given moment depending on the state of the cluster. The opportunistic resource pool increases its capacity as more fixed-size exclusive jobs finish their executions and release allocated GPUs, and decreases its capacity as incoming fixed-size jobs are allocated. To take advantage of this type of transient resource and circumvent the long queues associated with static allocations, high-throughput applications have traditionally been architected to operate on 
this elastic resource pool
to maximize the throughput over a long period of time. Furthermore, this elasticity helps increase the cluster-wide resource utilization as previously idle GPU cycles are now filled with high-throughput applications running opportunistically. Note that to accommodate this dynamicity, monolithic applications are typically decomposed into many smaller tasks which are independently executed on remote nodes or GPUs.

\textbf{Motivation}. The integration of lightweight LLMs, however, presents a fundamental challenge to this opportunistic paradigm. The volatile nature of the resource pool means that an executing task may be preempted at any time. For conventional HPC tasks, the cost of such a preemption is often manageable. For a lightweight LLM inference task, it is \textbf{catastrophic}. The startup procedure for a single task—loading a multi-billion parameter model from a distributed filesystem to local disk, then into host memory, and finally into GPU memory—is an I/O-bound process that can take minutes. Upon preemption, this entire costly procedure must be repeated from scratch on a new resource, and frequently dominates the task's useful computation time. While batching many inferences can amortize this cost, it simultaneously linearly increases the task's runtime, elevating the risk that the task will be preempted before completion and loses all its work. Thus, a user must be deliberate and careful in tuning the inference batch size per task to balance between the startup cost and the preemption rate. 

This tuning process unfortunately adds another layer of complexity for non-technical users, and the optimal batch size, which depends on the startup cost and the preemption rate, is not straightforward to derive either, even for technical users. While the startup cost can be profiled on a local GPU, a typical GPU cluster has many models of GPUs, reflecting the cluster's evolution over time as older GPUs are gradually phased out and newer GPUs are incrementally added in. Table \ref{tab:gpus} shows the heterogeneity of our local HPC cluster with 8 major GPU models spanning 8 years and accounting for 75\% of all GPUs in the cluster. This \textbf{heterogeneity} thus complicates the validity of an application's runtime profiling with a local GPU:  opportunistic GPUs can come with \textit{any} GPU model, and an optimal batch size for one GPU model doesn't necessarily translate to optimality on other GPUs. Furthermore, the rate of GPU preemption is completely \textit{dynamic and unpredictable over time} as it depends on, among other factors, the current load on the local cluster, the arbitrary resource demands from other GPU jobs, and the specific allocation and scheduling policies of the cluster manager.

All issues we discussed above form the following central research question of this paper: 
\textbf{Is there a way that we can transform existing high-throughput lightweight LLM inference applications such that they can execute efficiently on heterogeneous opportunistic GPU resources without repeatedly paying the startup cost upon preemption and/or incurring a high toll on HPC users?}

\textbf{Limitation of state-of-the-art approaches}.
Existing methodologies for managing elastic and fault-tolerant computations are not well-suited to resolve the core tension between the high startup cost of LLM inference and the volatile nature of opportunistic resources.

First, conventional autoscaling frameworks\cite{qiu2023aware,zou2024optscaler,augustyn2024tuning,catillo2023survey} are fundamentally mismatched with the opportunistic resource model. Autoscaling systems operate on a \textbf{proactive} principle: the application itself initiates scale-up or scale-down events based on its internal workload, such as a rising queue of user requests. In the opportunistic HPC setting, the application is purely \textbf{reactive}: it has no control over its allocated resources, and GPUs are allocated and preempted by the cluster manager based on external priorities. Therefore, an application cannot "request" more resources to meet demand, nor can it "release" them gracefully: it must simply adapt to the resources it is given, whenever they appear or disappear.

Second, traditional fault-tolerance techniques like progress checkpointing \cite{goulart2023checkpointing, islam2012mcrengine, siachamis2024checkmate} offer only a partial and inadequate solution. While checkpointing the results of completed inferences allows a task to protect its progress, it does not address the primary problem: the prohibitive model loading cost. Upon preemption, a new task must still be instantiated on a different GPU, incurring the full startup penalty before it can resume work from the last checkpoint.

\begin{table}[t]
    \centering
    \begin{tabular}{|c|c|c|}
        \hline
         Device Name &  Release Year & Count \\
        \hline
               NVIDIA Quadro RTX 6000 & 2018 & 106\\
        \hline
        NVIDIA A10 & 2021 & 78 \\
        \hline
        NVIDIA TITAN X (Pascal) & 2016 & 69 \\
        \hline
        NVIDIA GeForce GTX 1080 Ti & 2017 & 63 \\
        \hline
        NVIDIA RTX 6000 Ada Generation & 2022 & 36 \\
        \hline
        NVIDIA GeForce GTX TITAN X & 2015 & 34 \\
        \hline
        NVIDIA A40 & 2020 & 26 \\
        \hline
        NVIDIA H100 80GB HBM3 & 2023 & 15 \\
        \hline
    \end{tabular}
    \caption{8 Major GPU Models in the Local Cluster}
    \label{tab:gpus}
    \justifying
    \noindent
    \textit{Our local HPC cluster contains 567 GPUs in total, with 75\% of them in one of the 8 major models in the table, showing the heterogeneity of the available GPUs and emphasizing the challenges of running scientific applications opportunistically on heterogeneous resources.}
\end{table}

\textbf{Key insights and contributions}. Our approach is founded on a key insight: the primary performance bottleneck is not the computation but the tight coupling of inference execution with expensive context initialization. We propose to \textbf{decouple} these elements by elevating the LLM context (e.g., the model loaded in a GPU) to a first-class, persistent entity within the cluster, a technique we term Pervasive Context Management.

In this technique, contexts are deliberately held on remote nodes for reuse across multiple tasks rather than being torn down after each task's completion. When a task is preempted from one GPU, it is simply requeued and rapidly rescheduled to another opportunistic GPU that already holds the required context. Furthermore, when new GPUs join the resource pool, our system can bootstrap them by transferring an existing context template directly from another node, cutting down data transfer time and preventing a bottleneck at the shared filesystem. This strategy effectively makes the high cost of context initialization a one-time, amortizable expense. It also alleviates the complex problem of searching for an optimal batch size as the startup cost is now shared across many independent tasks. Figure 1 illustrates the conceptual differences between the static allocation model, the standard opportunistic model, and our proposed model that combines opportunistic resources with Pervasive Context Management.

Based on these ideas, this paper makes the following contributions:
\begin{enumerate}
    \item We implemented a high-throughput context-agnostic fact verification application backed by lightweight LLMs in the Parsl-TaskVine distributed data-intensive framework\cite{babuji2019parsl, sly2023taskvine, phung2023maximizing}.
    \item We detailed a quick application transformation that makes the application context-aware via the Pervasive Context Management technique.
    \item We conducted a comprehensive evaluation demonstrating that this transformation, enabled by Pervasive Context Management, significantly reduces the end-to-end execution time of the application by up to 72.1\% (from 3 hours to 48 minutes), and allows the application to scale opportunistically on up to 32.8\% of all GPUs in the cluster and further reduces the execution time to 13 minutes.
\end{enumerate}

\textbf{Limitation of the proposed approach}. The primary constraint of the proposed approach is that our Pervasive Context Management system is designed for lightweight LLMs (up to billions of parameters depending on the GPU setup per node) whose context can fit within the resources of a single compute node. This is a direct consequence of the nature of opportunistic resources in HPC clusters, which are typically allocated and preempted on a per-node basis. Additionally, our system introduces its own management overhead for replicating and caching contexts, and its effectiveness is contingent on this overhead remaining substantially lower than the cost of repeated cold starts from a shared filesystem.

\section{Implementation of a Context-Agnostic Fact Verification Application}

\begin{figure}[t]
    \includegraphics[width=\columnwidth]{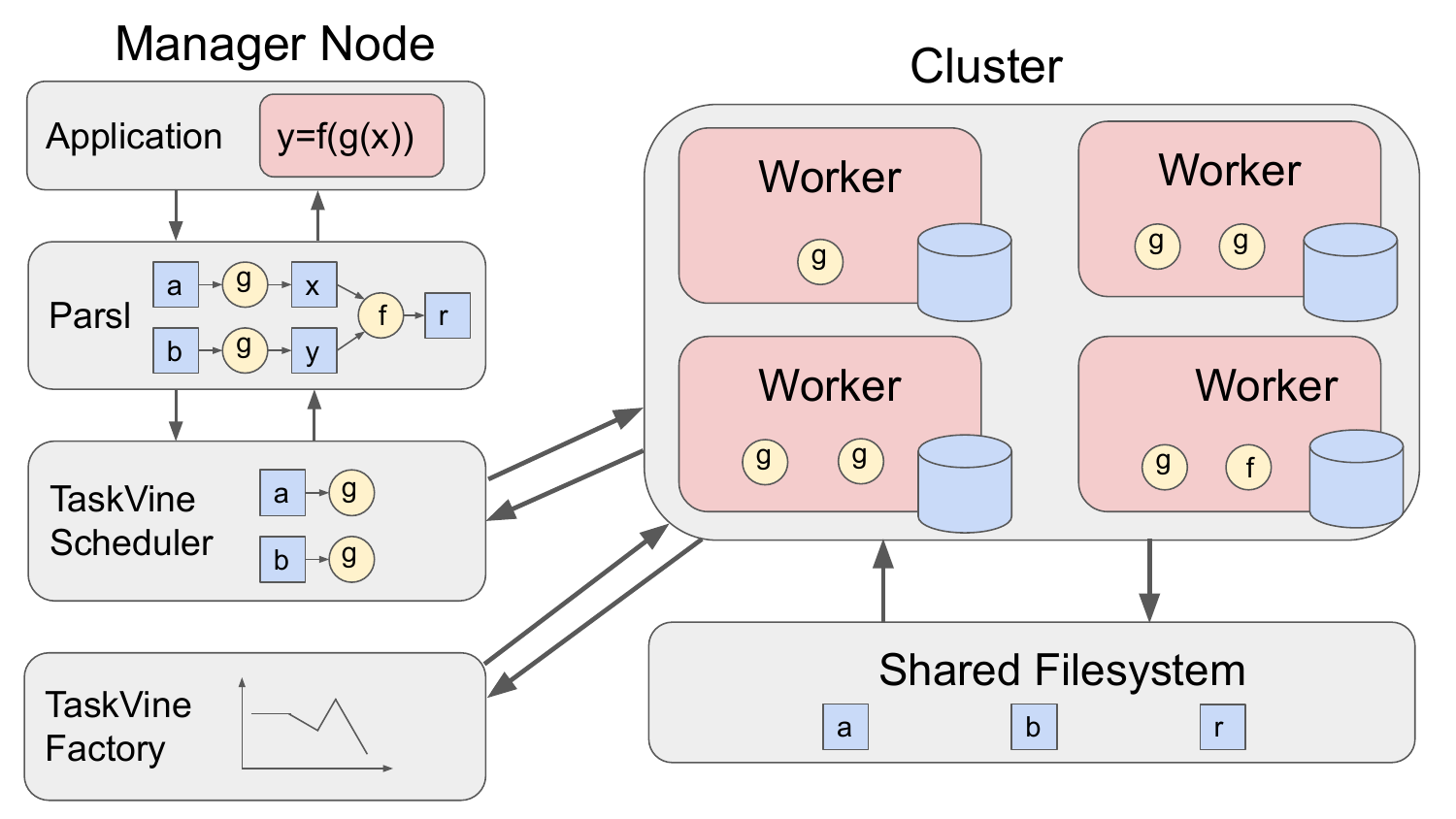}
    \caption{Overview of the Parsl-TaskVine Framework}
    \justifying
    \noindent \textit{The application defines LLM computations via Python functions and passes them to Parsl. Parsl manages dependencies between functions and sends ready ones to the TaskVine scheduler. The scheduler manages resources on workers, schedules functions to available ones, and controls their execution and I/O patterns. The TaskVine factory monitors the opportunistic resources and adjusts the quantity of workers accordingly.}
    \label{fig:software-stack}
\end{figure}

\textbf{Overview of the Parsl-TaskVine framework.}
The underlying framework that powers the fact verification application is the integrated software stack of two dynamic workflow systems - Parsl\cite{babuji2019parsl} and TaskVine\cite{sly2023taskvine}. Parsl is a Python-native parallel library that allows users to express their computational needs via generic Python functions and automatically scales the computation on thousands of compute nodes, mainly focusing on flexibility, portability, and ease of use. 
TaskVine is a low-level data-intensive workflow execution engine that allows users to express low-level details about tasks and their inter-relationships. It then extracts values from the provided information to make intelligent scheduling and optimization decisions that accelerate large-scale data processing applications\cite{phung2024accelerating, phung2024adaptive}.

Figure \ref{fig:software-stack} shows how these two workflow systems work together in the big picture.
On the manager node, a user expresses their application's computational needs (e.g., LLM inferences) via generic Python functions. Once the application is run and these functions are invoked, they are intercepted and passed to Parsl for inter-function dependency management and function-to-task translation. Parsl sends ready tasks to the TaskVine scheduler, where they are examined for common execution and I/O patterns and scheduled for execution on workers accordingly. 
The TaskVine scheduler manages resources in the system via TaskVine workers, where each worker is a small standalone pilot job that waits for instructions from the TaskVine scheduler and operates duly. Once tasks are completed, workers communicate the results back to the scheduler, which forwards them back to the application level. 
Note that the TaskVine scheduler does not delegate the local resource management to individual workers: each task comes with a specific amount of resource allocation, and each worker is directed by the TaskVine scheduler on how to utilize any local resource type (CPU, memory, SSD, GPU). The pool of resources is maintained by the TaskVine factory, a daemon-like process that monitors the current resource pool and adjusts it based on a given resource policy and the current load of the cluster.

\begin{figure}[t]
\begin{minted}[fontsize=\small, linenos]{python}
from parsl import python_app
@python_app
def infer(model_path, claims):
    ...
    model = AutoModel.from_pretrained(model_path).to('gpu')
    verdicts = [model.generate(claim) for claim in claims]
    return verdicts
model_path = ...
claims = ...
verdicts = infer(model_path, claims).result()
\end{minted}
\caption{Code Example of a Context-agnostic Fact Verification Application}
\label{fig:sample-code-context-agnostic}
\justifying
\noindent
\textit{An inference function is annotated with a Parsl-provided decorator, and remote execution is triggered by invoking the function as usual. Calling the ".result()" method blocks the execution until the result of the invocation is returned.}
\end{figure}

\textbf{Implementing the context-agnostic fact verification application.}
Given this software stack, it is then straightforward for a user to implement a high-throughput lightweight LLM inference application. 
A user first defines an arbitrary computation involving LLM inferences in a Python function. This function then flows through Parsl and the TaskVine scheduler to a TaskVine worker as a task to be executed. 
Each worker is allocated with a small number of GPUs such that a given task can run comfortably. Once the task completes, inference results are sent from the worker back to the application as described above. The scheduler has a queue of ready tasks, and its main job is to occupy connected workers with tasks at any given time. Therefore, the application will make progress as long as there are workers connected to the scheduler.

Additionally, this software stack provides a seamless integration with opportunistic resources. The scheduler on the manager node directs all workers on what to do and thus keeps a globally consistent view of the application. This means that workers can leave and join the pool freely as tracked by the TaskVine scheduler and adjusted by the TaskVine factory, and any preempted task is detected, retrieved, and re-inserted into the queue of ready tasks by the scheduler.

Figure \ref{fig:sample-code-context-agnostic} shows a simplified implementation of an inference function of the fact verification application (we delay the full description of this application to Section \ref{subsec:settings}). To minimize the toll on non-technical users when scaling up a local application, Parsl provides a decorator ("@python\_app") that wraps around a typical inference function. Remote execution is triggered when the user invokes the inference function as usual, allowing Parsl to intercept the inference invocation and prepare it for remote execution. The inference is then given to TaskVine for scheduling and execution, and the result is transparently sent back to the application. Alternatively, a user can instruct the local execution to wait for the result by calling the ".result()" method before continuing its execution flow.

Notice that this context-agnostic implementation forces a function invocation to reload its context whenever it is preempted from an opportunistic resource as it couples the expensive context initialization with the actual inference execution into one executable unit (i.e., a task). The next section shows how we can decouple these two computational elements to allow the efficient reuse of a one-time startup cost per unit of opportunistic resources over multiple inference tasks. 

\begin{figure}[t]
    \includegraphics[width=\columnwidth]{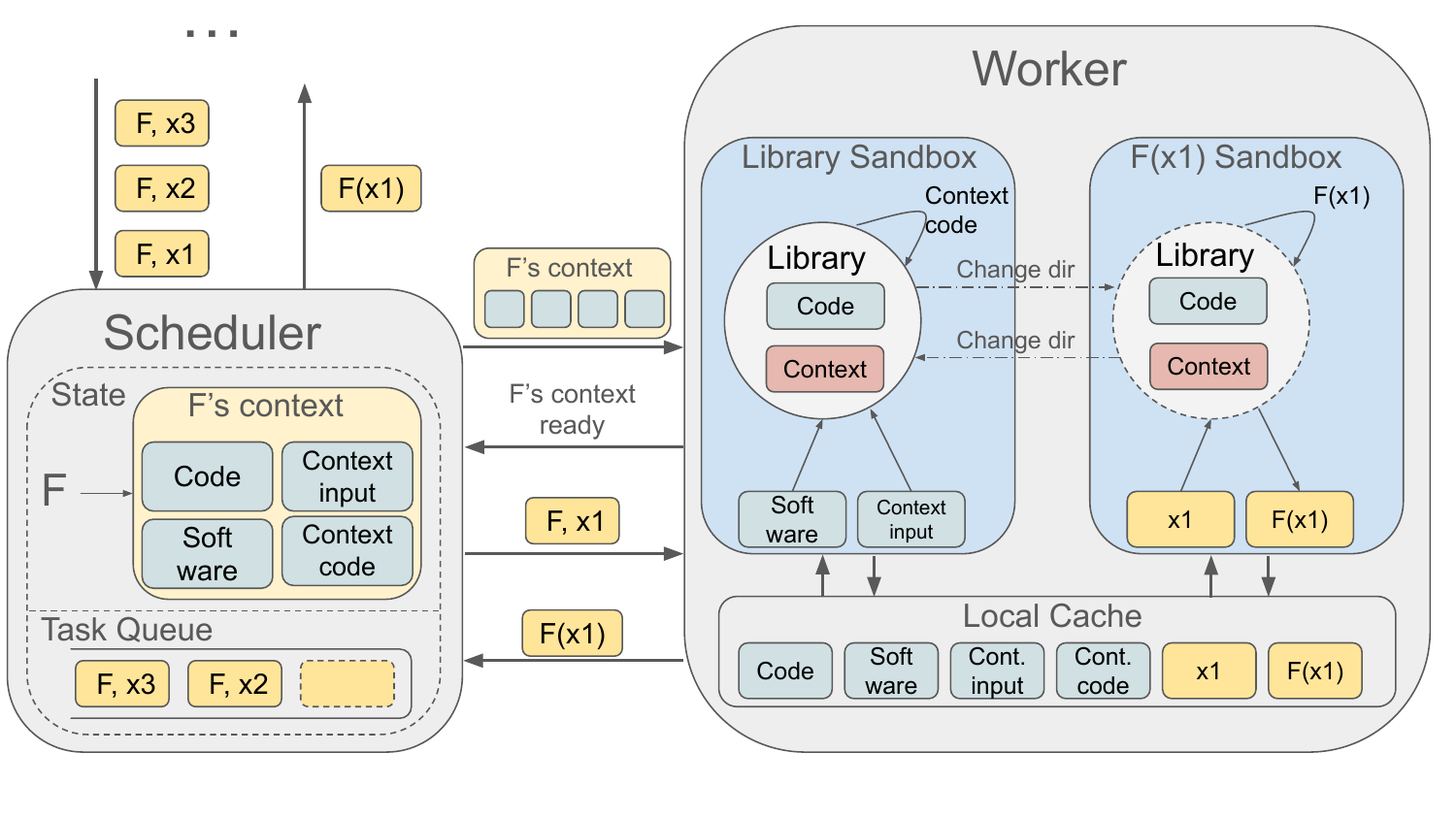}
    \caption{Overview of Pervasive Context Management}
    \justifying
    \noindent \textit{The TaskVine scheduler analyzes F for its context upon the first invocation request, and sends it to the worker. The library process in the worker registers F's code and creates and holds an instance of F's context. This context and registered code are then used to execute the request, and subsequent requests reuse this existing context to speed up their executions.}
    \label{fig:detailed-context-mgmt}
\end{figure}

\section{Transforming the Application to Enable Pervasive Context Management}
\textbf{Overview of the Pervasive Context Management technique.} Figure \ref{fig:detailed-context-mgmt} gives a general example of how a function context is retained and reused in the Parsl-TaskVine stack. An application (not shown) starts up and invokes function F three times with arguments x1, x2, and x3, respectively. Parsl (not shown)  sees that these three functions don't have any dependency between them and converts them into ready tasks to be sent to the TaskVine scheduler. The scheduler examines F and discovers its context recipe, including F's code, software dependencies, context code, and context inputs, (we discuss the context transformation later in this section) and sends this context recipe to be materialized and  hosted on a given worker as part of F(x1) execution. The worker, upon receiving the context recipe, stores all of its components in a local cache and fork-execs a special process called "Library". The Library process is responsible for materializing and hosting F's context from its recipe and will cooperate with the worker to execute subsequent invocations of F. 
Upon the stage-in of F's context into its sandbox, the library registers F's code, executes the context code, stores the resulting context as an internal state in its process, and lets the worker know it's ready for invocations of F. The scheduler, upon receiving 
this ack from the worker, sends the first invocation request of (F, x1). 
The worker stores x1 in its cache, creates a sandbox for the invocation, and pings the library. The library then changes its working directory to F(x1)'s sandbox and executes the invocation directly in its address space, which already contains F's context, before returning to its sandbox. The result of the invocation is then returned to scheduler, which marks the completion of F(x1) and forwards the result to the application. Executions of (F, x2) and (F, x3) then reuse F's context via the library and follow the same path (F, x1) took.

\begin{figure}[t]
\begin{minted}[fontsize=\small, linenos]{python}
from parsl import python_app
def load_model(model_path):
    ...
    model = AutoModel.from_pretrained(model_path).to('gpu')
    return {'model': model}
@python_app
def infer_model(claims, parsl_spec):
    from parsl import load_variable_from_serverless
    model = load_variable_from_serverless('model')
    verdicts = [model.generate(claim) for claim in claims]
    return verdicts
model_path = ...
claims = ...
parsl_spec = {'context': [load_model, [model_path], {}]}
verdicts = infer_model(claims, parsl_spec).result()
\end{minted}
\caption{Code Example of a Context-aware LLM Inference Application}
\label{fig:sample-code-context-aware}
\justifying
\noindent
\textit{The previous inference function is now broken down into two new functions: "load\_model" that creates the model state in the GPU, and "infer\_model" that reuses the existing model state and runs inferences. "infer\_model" specifies its model context as an argument, and the remote execution is triggered as usual.}
\end{figure}

\textbf{Code transformation to enable Pervasive Context Management.}
It is then effortless to see how The Pervasive Context Management technique benefits the new class of high-throughput lightweight LLM inference applications on opportunistic resources. Since the startup cost of initializing an LLM is expensive, an application should define it as a context to the actual inference task. When newly available opportunistic resources arrive to the resource pool, the manager sends the context template to be initialized once per node and retained for subsequent reuses. As multiple inference tasks arrive to the task queue, the manager sends them to nodes that already have the context initialized for immediate inference execution. When resources are preempted by the cluster, these tasks are seamlessly requeued by the manager for execution on other nodes that already host the needed context, eliminating the need to reinitialize the LLM from scratch per task.

Figure \ref{fig:sample-code-context-aware} shows a 
code example of how the fact verification application can be quickly transformed to benefit from this technique. We first decouple, or split, the previous inference function into two new functions: "load\_model" and "infer\_model". Lines 2-5 define the "load\_model" function that creates an LLM context by loading its parameters from disk to GPU and returns this context via a dictionary to the Library process. This dictionary informs the Library of the relevant context to later be exposed to the actual invocation. Lines 7-12 define the actual computation via the "infer\_model" function that loads the model directly from the context held by the library (instead of loading it from scratch), executes the inferences, and returns the results. Lines 14-17 connect the missing pieces of the example where the context computation is defined via the \verb|parsl_spec| variable, and "infer\_model" brings this context reference along with its inputs to the scheduler for execution.

\section{Evaluation}
\label{sec:eval}

This section begins with an in-depth description of the fact verification application along with the general experiment settings that apply to all evaluation efforts. Our evaluation then aims to answer the following research questions:
\begin{itemize}
    \item \textbf{RQ1 - Application Performance on Static Resources}. How well does the application perform with and without the Pervasive Context Management technique on statically allocated resources?
    \item \textbf{RQ2 - Application Sensitivity to Varying Inference Batch Sizes}. How does enabling Pervasive Context Management help users pick the right inference batch size for the application?
    \item \textbf{RQ3 - Application Performance with Aggressive Resource Preemption}. How well does the context-aware application handle aggressive resource preemption from the cluster manager?
    \item \textbf{RQ4 - Application Performance with Opportunistic Resources}. How well does the context-aware application scale opportunistically when the capacity of transiently available resources in the cluster fluctuates?

\end{itemize}

\subsection{Experiment Settings}
\label{subsec:settings}
\textbf{Application.}
Fact verification is an active area of research given the lightning rise of online mis- and dis-information\cite{zhang2023towards, gunjal2024molecular}.
Our application, Prompt for Fact (PfF), aims to find the \textit{optimal} prompt for a given LLM where it is used as a fact verifier to check the correctness of an arbitrary claim. 
Specifically, we use the training data from FEVER\cite{Thorne18Fever} as our dataset containing 145,449 claims, each of which is labeled with either \verb|SUPPORTED|, \verb|REFUTED|, or \verb|NOT ENOUGH INFO|. Each claim contains a statement about a given subject and a list of references to relevant Wikipedia pages.
Per the LLM, we use the recently released SmolLM2 model with 1.7 billion parameters\cite{allal2025smollm2}.
Our application takes the LLM and a prompt template, runs a full inference sweep across the dataset, and returns the fact verification accuracy. Note that the Pervasive Context Management technique is \textit{not} limited to a specific model or application and is applicable to all high-throughput lightweight LLM inference applications subject to the limitation as described in Section \ref{sec:intro}.

\textbf{Local cluster.}
Our local cluster runs two resource managers: Altair Grid Engine (AGE)\cite{age} as the main static batch manager, and  HTCondor\cite{thain2005distributed} as a resource backfiller on AGE's unallocated machines. There are 567 GPUs in total in the cluster with 18 different models (see Table \ref{tab:gpus} for 8 major GPU models). We run all experiments through the HTCondor resource manager. Our cluster provides access to data via the Panasas ActiveStor 16\cite{shaffer2017taming,panasas} shared filesystem with 77 nodes and supports up to 84 Gbs/s read bandwidth and 94k read IOPS. 

\textbf{Parsl-TaskVine framework}. We configure parameters of our Parsl-TaskVine software stack as follows. We enable the peer transfer feature that allows workers to communicate and send arbitrary data between each other, which allows workers to send context templates in a peer-to-peer fashion and bootstrap newly arrived workers. Each task's resource allocation includes 2 cores, 10 GBs of memory, 20 GB of disk, and 1 GPU, providing a comfortable amount of resources for a smooth inference execution. Each TaskVine worker has 2 cores, 10 GBs of memory, 70 GBs of disk, and 1 GPU, thus providing the worker with just enough resources to run tasks in a 1-to-1 manner to preserve claimed opportunistic resources and plenty of disk space for local caching.

\begin{figure}[t]
    \centering
    \includegraphics[width=\columnwidth]{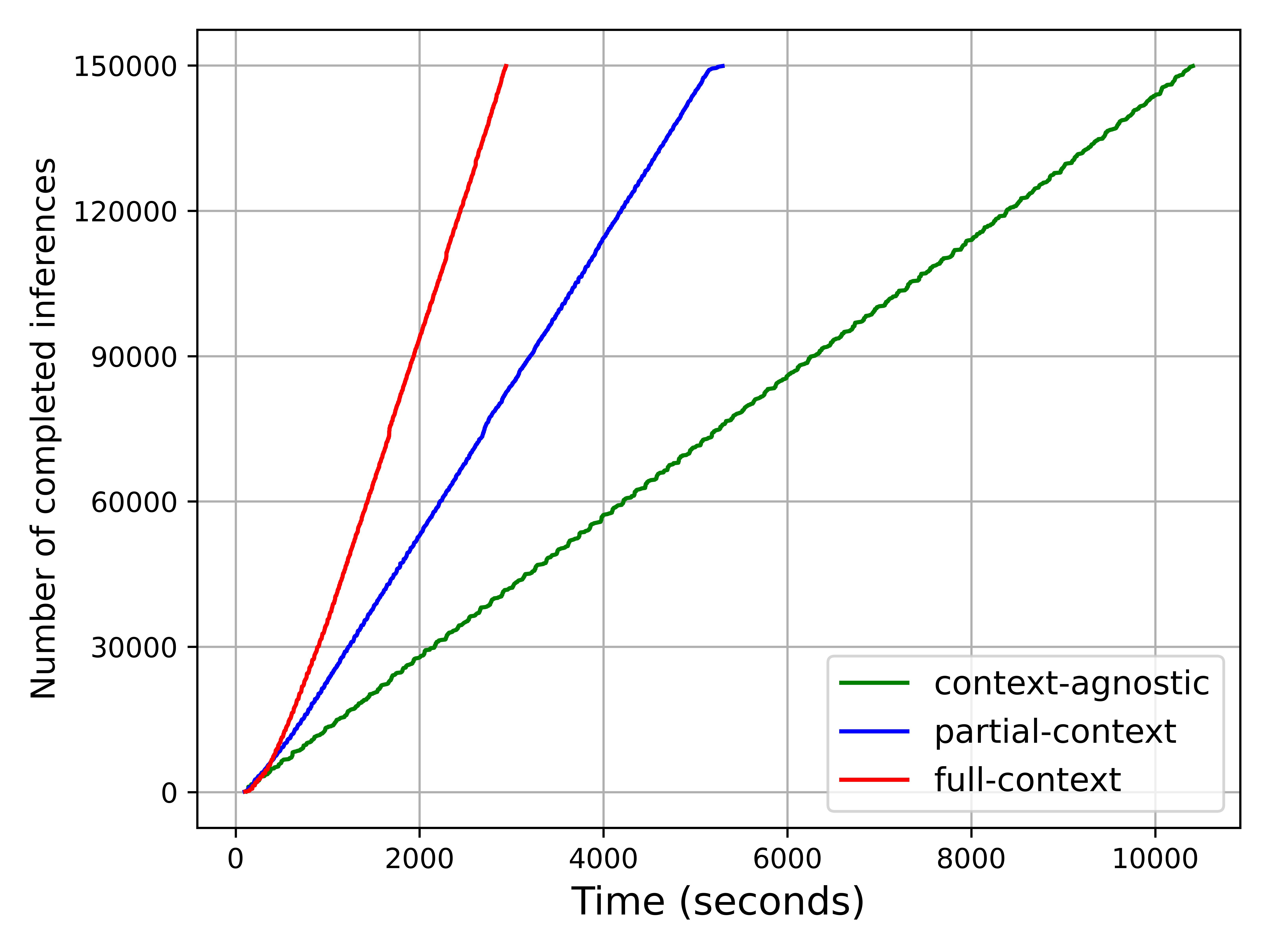}
    \caption{Execution Time of the Prompt for Fact Application with Increasing Level of Context Awareness on Static Resources}
    \label{fig:3_runtimes}
    \justifying
    \noindent
    \textit{The application is run with 3 levels of context awareness: context-agnostic (no context is encoded in the application, forcing LLM state initialization per task), partial-context (LLM context is persisted only on local disk, which includes GBs of model parameters and software dependencies), and full-context (the context further includes the state of the LLM model loaded in the local GPU for quick reuse). The end-to-end execution time of the application reduces drastically as it is run with a higher level of context awareness.}
\end{figure}

Finally, almost all experiments start with the same resource pool configuration consisting of 20 GPUs, where half are NVIDIA A10 and the other half are NVIDIA TITAN X (Pascal). This approach allows us to not only establish consistency and stability to our measurements and results but also mimic the heterogeneity of the actual GPU cluster.
This constraint is removed at the end which allows the application to have access to up to
186 opportunistic GPUs.
Storage-wise, the LLM takes up 3.7 GBs of disk and around 7.4 GBs of memory when fully loaded. The application's software dependencies are managed in a Conda\cite{conda} environment, containing 308 packages and totalling 10.5 GBs of disk.

\subsection{RQ1 - Application Performance with Static Resources}
\label{subsec:rq1}
To quantify the impact of the Pervasive Context Management technique on the Prompt for Fact application, we implement the \textit{context-agnostic} version of the application and transform it into two other versions: \textit{partial-context} and \textit{full-context}. We detail the differences between these versions as below:
\begin{itemize}
    \item \textbf{Context-agnostic}. In the \textit{context-agnostic} version, a task makes all I/O calls for the model parameters and software dependencies to the shared filesystem and creates a fresh state of the LLM model from scratch, involving moving GBs of data from the shared filesystem to the local disk of the execution node, to the node's memory, and finally to the node's GPU.
    \item \textbf{Partial-context}. Instead of transferring GBs of input data per task, \textit{partial-context} caches these common input data on the local disk of the remote nodes so that subsequent tasks that run inferences on the same model can reuse the data available locally instead of making remote I/O calls to the shared filesystem. Each task still has to create a fresh state of the LLM model in the GPU however.
    \item \textbf{Full-context}. This version fully unlocks the benefits of the Pervasive Context Management technique and caches the LLM model state in the GPU for efficient context reuse and fast context transfer between successive task executions. In other words, the LLM model state is created and loaded to the GPU once over a worker's lifetime.
\end{itemize}
All applications are run on statically allocated resources to isolate the results from other noises (e.g., fluctuations in capacity of opportunistic resources). Each task carries 100 inferences (i.e., inference batch size of 100), resulting in 1,500 tasks per application execution. 

Figure \ref{fig:3_runtimes} shows the end-to-end execution time of 3 versions of the Prompt for Fact application, each totaling 150,000 inferences. The result aligns with our expectation as the version with more context awareness has a significantly lower execution time. Specifically, the \textit{context-agnostic} version runs the longest, over 10.4k seconds, as each task has to repeatedly make remote I/O requests to load GBs common input data from the shared filesystem and construct a new GPU state of the LLM model upon initialization. \textit{Partial-context} instead caches GBs of common input data on local disk of remote nodes and effectively converts most remote I/O requests into local ones. This thus helps bring the execution time down to 5.3k seconds, a significant reduction of the end-to-end execution time of 49.1\%. \textit{Full-context} cuts down the execution time even further to 2.9k seconds, a reduction in execution time of 72.1\% and 45.3\% compared to \textit{context-agnostic} and \textit{partial-context}, respectively. This is because it eliminates both a bulk of unnecessary remote I/O requests for common input data and the need for a task to reconstruct the LLM state in a GPU upon startup as a task can now reuse an available context already initialized in a worker. 

We then conclude that transforming the Prompt for Fact application to enable the Pervasive Context Management technique allows a considerably more efficient execution with a huge reduction in the end-to-end execution time on static resources.

\begin{figure}[t]
    \centering
    \includegraphics[width=\columnwidth]{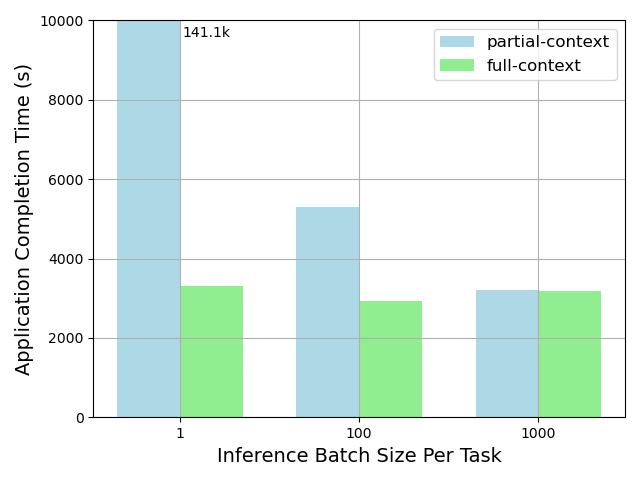}
    \caption{Effect of Inference Batch Size to the Application's Execution Time}
    \label{fig:batch-size}
    \justifying
    \noindent
    \textit{Partial-context and full-context are run with 3 different batch sizes: 1, 100, and 1000. Partial-context introduces a large variance of execution time across batch sizes due to the expensive LLM model initialization cost per task, even with cached common input data. On the other hand, full-context stabilizes this range of execution time as each model is initialized once per GPU and reused across multiple tasks instead of once per task, alleviating the effect of a wrong choice of inference batch size.}
\end{figure}

\begin{figure}[t]
    \centering
    \includegraphics[width=\columnwidth]{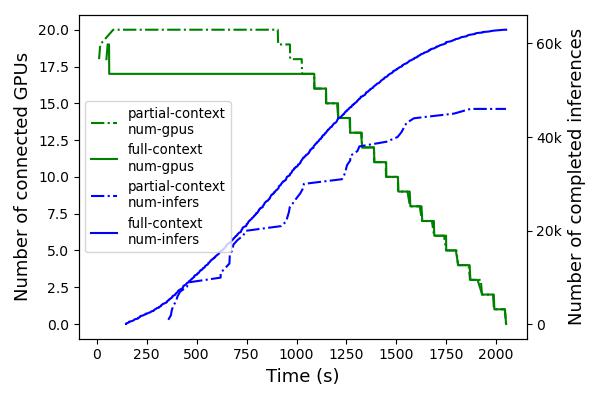}
    \caption{Number of Completed Inferences with Aggressive Resource Preemption}
    \label{fig:worker-preemption}
    \justifying
    \noindent
    \textit{This figure shows the number of completed inferences over time between partial-context and full-context with aggressive resource preemption from the cluster (1 GPU preemption per minute). Despite an early drop of 3 GPUs, full-context still completes 16.9k inferences more than partial-context, and consistently has a higher inference completion rate at any given time.}
\end{figure}

\subsection{RQ2 - Application Sensitivity to Varying Inference Batch Sizes}
Subsection \ref{subsec:rq1} demonstrates how the Pervasive Context Management technique enables an efficient execution of the Prompt for Fact application and addresses the first part of the central research question as posed in Section \ref{sec:intro}. This subsection then addresses the second part of the question and shows how this technique makes it easy for users to pick an inference batch size without worrying about optimal and/or sub-optimal application execution.

Figure \ref{fig:batch-size} shows the execution time of \textit{partial-context} and \textit{full-context} with 3 batch sizes: 1, 100, and 1000 (we skip \textit{context-agnostic} as it is clearly suboptimal as demonstrated in Subsection \ref{subsec:rq1}.) With a wrong inference batch size of 1, \textit{partial-context} takes a disastrous hit in its performance and needs 141.1k seconds to complete end-to-end. This is because each inference now needs to load the model from scratch, and the overhead of the model initialization dominates the total execution time. Even at the batch size of 100, \textit{partial-context} still takes 5.3k seconds and is still far from the best execution time with a batch size of 1000 at 3.2k seconds.

On the other hand, \textit{full-context} allows a much more stable range of execution time across all batch sizes. The application runs the "worst" with an inference batch size of 1 at 3.3k seconds, and the "best" with batch size of 100 at 2.9k seconds. The range of execution time is thus limited to approximately 400 seconds, or 13.6\% of the best execution time, over the range of possible batch sizes of 1000 (from 1 to 1000). Such a stable range of execution time comes from the context cache and reuse between tasks, and the performance difference is only the cumulative overhead of the context transfer that happens once per task. Thus, enabling the Pervasive Context Management technique ensures that the application can never run disastrously with a wrong batch size and its execution is always optimal or near-optimal, removing the worry of  batch size tuning from users.

\begin{figure}[t]
\begin{subfigure}[t]{0.45\columnwidth}
\includegraphics[scale=0.24]{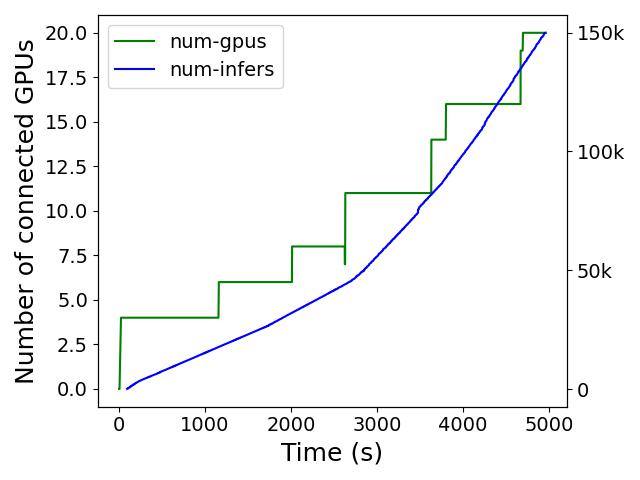} 
  \captionsetup{justification=centering}
  \caption{Low opportunistic capacity}
  \label{plt:low}
 \end{subfigure}
 \begin{subfigure}[t]{0.45\columnwidth} 
 \includegraphics[scale=0.24]{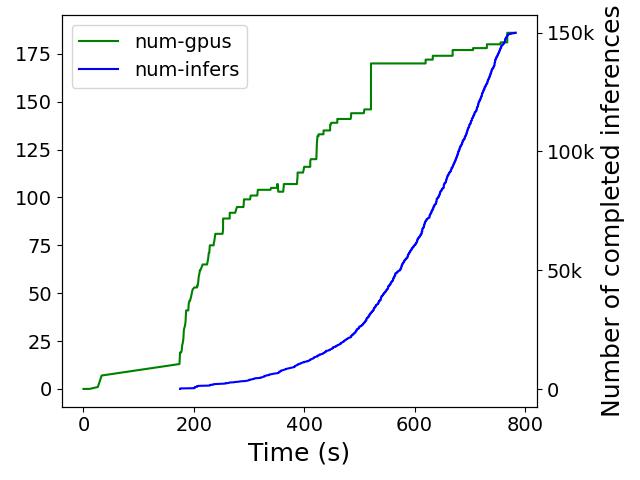} 
 \captionsetup{justification=centering}
  \caption{High opportunistic capacity}
  \label{plt:high}
 \end{subfigure}

\caption{Application Scaling on Opportunistic Resources over Time}
\label{plt:free-runs}
\justifying
\noindent
\textit{The Prompt for Fact application scales opportunistically on the local cluster over time at low opportunistic capacity (left) and high opportunistic capacity (right). The application scales linearly with the constantly changing amount of available resources.}
\end{figure}

\subsection{RQ3 - Application Performance with Aggressive Resource Preemption}

Opportunistic resources fluctuate frequently, increasing the capacity as more jobs exit and release their previously claimed resources, and decreasing the capacity as new jobs are allocated and scheduled by the cluster manager. This subsection focuses on the latter and quantifies how well \textit{full-context} executes the application efficiently with aggressive resource preemption.

Figure \ref{fig:worker-preemption} shows the scenario where resources are preempted aggressively from the application by the cluster manager with the preemption rate of 1 GPU per minute from the 900-second mark until the opportunistic resource pool of the application is depleted (we preempt all NVIDIA A10s before NVIDIA Titan X Pascals). For \textit{partial-context}, we can see the "rugged" rate of inference completion that gradually flats out from the opportunistic resource depletion and ends with 46k completed inferences. This is due to the expensive LLM model initialization cost per task that blocks goodput until the LLM model is fully loaded in a GPU, creating the effect of low goodput when the LLM models are being loaded on remote workers, and high goodput when inferences are executed.

Per \textit{full-context}, despite having an early drop of 3 GPUs due to external preemption from the cluster manager, \textit{full-context} still completes 16.9k inferences more than \textit{partial-context} (ends with 62.9k completed inferences), and consistently has a higher inference rate than \textit{partial-context}. Notice the smooth inference completion rate as it shows how tasks from preempted GPUs are seamlessly requeued and rerun with an already GPU-initialized LLM context, removing the rugged-like effect on the application's inference rate and allowing the slope to smoothly flat out at the end compared to that of \textit{partial-context}. Thus, the Pervasive Context Management technique helps the application smoothly make more progress even when resources are preempted aggressively.

\subsection{RQ4 - Application Performance with Opportunistic Resources}
Finally, we focus on how well the context-aware application (i.e., \textit{full-context}) scales opportunistically in the local cluster. Figure \ref{plt:free-runs} shows the number of opportunistic GPUs used and the number of completed inferences over time for \textit{full-context} when the cluster has low ( Subfigure \ref{plt:low}) and high (Subfigure \ref{plt:high}) opportunistic capacity. In Subfigure \ref{plt:low}, the application only starts out with 4 GPUs and gradually goes to 20 opportunistic GPUs, finishing around 5000 seconds. Note that even with a limited amount of GPUs, the application still makes consistent progress as the completed inference rate scales linearly with the amount of opportunistically connected GPUs. On the other hand, when the cluster has many jobs exiting and releasing their claimed GPUs (Subfigure \ref{plt:high}), the application quickly grabs up to 186 transiently available GPUs (32.8\% of all GPUs in the cluster) and finishes the execution in only 783 seconds. This thus shows that the Pervasive Context Management technique allows the Prompt for Fact application to swiftly react and scale to the ever-changing amount of opportunistic GPUs in the cluster over time.

\section{Related Works}
\label{sec:related}
\textbf{Spot Instances.} The use of underutilized compute capacity is a well-established practice in both commercial cloud computing and High-Performance Computing (HPC), though the implementation and guarantees vary. Cloud providers offer discounted Spot or Preemptible Instances\cite{GoogleCloudSpotVMs, AWSEC2Spot, AzureSpotVMs} that can be reclaimed at any time, similar to how opportunistic resources are used in HPC. While several studies\cite{mao2025skyserve, miao2024spotserve} have explored running stateful applications like LLM inference on these cloud resources, they rely on a critical feature: a preemption warning\cite{AWSSpotTermination,GoogleCloudSpot,AzureSpotVMsDoc}. This notification period, typically 30 to 120 seconds, allows applications to checkpoint state or transfer work before termination. On the contrary, opportunistic resources in many HPC environments offer no such warning, and preemption is instantaneous, rendering traditional state-saving mechanisms ineffective. Our work addresses this specific challenge by introducing the Pervasive Context Management technique designed to handle the abrupt and unpredictable nature of preemption in HPC clusters by retaining the common computational context between inferences in all connected GPUs.

\textbf{LLM Inference Optimization.}
Many works optimize the inference process of huge LLMs by outputting several tokens in one forward pass based on the speculative decoding scheme\cite{leviathan2023fast,spector2023accelerating, chen2024cascade, svirschevski2024specexec}. This scheme assumes that an LLM generating tokens sequentially takes too much time and resources, especially with easy-to-predict tokens. To speed this up, a smaller LLM is used to predict the next K tokens in advance, and the original LLM can make one forward pass that accepts tokens it agrees with and rejects others instead of making K forward passes.
Other works focus on KV cache and memory management on both a single GPU and a pool of GPUs. Kwon et. al.\cite{kwon2023efficient} introduce a virtual paging mechanism that divides the dynamically-sized KV cache into blocks to remove GPU memory fragmentation, while Lin et. al.
\cite{lin2024infinite} distribute the KV cache and the attention computation to many GPUs.
Cloud deployment of inference serving is also an active area of research. Fu et. al.\cite{fu2024serverlessllm} use local storage of individual instances to cache and distribute model checkpoints among each other. Our work extends the usage of local storage to memory and GPUs to hold and distribute the computational context.

\textbf{Workflow Systems.}
Workflow systems evolve from the traditional resource managers and allow applications to express complex relationships between tasks via a directed acyclic graph (DAG) instead of a bag of tasks\cite{deelman2015pegasus, turilli2019middleware, zheng2017deploying, phung2021not}. These systems typically focus on applications' reliability, performance, and portability via novel architectural designs and runtime optimizations, but require users to describe the computational needs in detail via complicated and non-uniform abstractions. More modern workflow systems\cite{babuji2019parsl, rocklin2015dask, moritz2018ray} tackle this usability problem by providing Pythonic abstractions that enable users to wrap their computational needs neatly into Python functions and translating these functions into tasks deployable to remote nodes. Our Parsl-TaskVine integration follows this movement and allows users to easily describe their computations in Python without losing performance, reliability, or portability. 
The Parsl-TaskVine stack extends this movement one step further with the support of computational context sharing between tasks on contrary to the traditional view of complete inter-task independence.

\section{Conclusion}
The LLM technology has been improving at an astonishing rate over the years and promises an unprecedented leap in human productivity. The current infrastructure however is ill-equipped to deal with the massive spike in interests and computational demands from LLM scientists and practitioners and requires new approaches to efficient resource management. This paper shows that, through the Pervasive Context Management technique, the new class of high-throughput lightweight LLM inference applications can be run efficiently on heterogeneous opportunistic GPU clusters without either sacrificing performance or imposing a heavy complexity toll on HPC users.

\bibliographystyle{ACM-Reference-Format}
\bibliography{main}

\end{document}